\documentclass[a4paper,noarxiv,twocolumn]{quantumarticle}
\usepackage[utf8]{inputenc}

\usepackage[numbers,sort&compress]{natbib}
\usepackage{hyperref}
\usepackage{amsmath}
\usepackage{amssymb, dsfont}
\usepackage{cleveref}
\usepackage{braket}
\usepackage{float}
\makeatletter
\let\newfloat\newfloat@ltx
\makeatother

\usepackage{algorithm}
\usepackage[noend]{algpseudocode}
\algnewcommand{\LeftComment}[1]{\State \(\triangleright\) #1}
\usepackage{slashbox}

\usepackage{booktabs}

\title{Belief propagation as a partial decoder}
\author{Laura Caune$^*$} 
\affiliation{Riverlane, Cambridge, CB2 3BZ, UK}
\author{Brendan Reid} 
\affiliation{Riverlane, Cambridge, CB2 3BZ, UK}
\author{Joan Camps} 
\affiliation{Riverlane, Cambridge, CB2 3BZ, UK}
\author{Earl Campbell}
\affiliation{Riverlane, Cambridge, CB2 3BZ, UK} 
\affiliation{Department of Physics and Astronomy, University of Sheffield, Sheffield S3 7RH, UK}

\date{June 2023}
\begin{document}

\maketitle
\def\thefootnote{*}\footnotetext{laura.caune@riverlane.com}\def\thefootnote{\arabic{footnote}}

\begin{abstract}
    One of the fundamental challenges in enabling fault-tolerant quantum computation is realising fast enough quantum decoders. We present a new two-stage decoder that accelerates the decoding cycle and boosts 
  accuracy.  In the first stage, a partial decoder based on belief propagation is used to correct errors that occurred with high probability.  In the second stage, a conventional decoder corrects any remaining errors.  We study the performance of our two-stage decoder with simulations using the surface code under circuit-level noise.  When the conventional decoder is minimum-weight perfect matching, adding the partial decoder decreases bandwidth requirements, increases speed and improves logical accuracy.  Specifically, we observe partial decoding consistently speeds up the minimum-weight perfect matching stage by between $2$x-$4$x on average depending on the parameter regime, and raises the threshold from $0.94\%$ to $1.02\%$.
\end{abstract}

\section{Introduction}

Quantum computers are expected to disrupt domain areas where quantum algorithms are much faster than their classical counterparts. These algorithms typically require running deep quantum circuits. For the output of these circuits to be meaningful, one needs minimal corruption from errors that arise due to the difficulty of isolating and controlling quantum systems.

The goal of quantum error correction (QEC) is to reduce the effect of noise within a quantum computer: building in redundancies to protect fragile quantum systems. This is achieved with QEC codes that encode a small number of (logical) qubits into a large array of (physical) qubits. If the error rate of these physical qubits is below a certain threshold, the logical qubit will exhibit a significantly reduced effective error rate. Simply put, the logical qubit will outperform the sum of its parts. We invite the reader to inspect Refs.~\cite{Dennis_2002,campbell2017roads,Roffe_2019_QEC} (and references therein) for a more complete and pedagogical overview of QEC.

In this work we focus on the dual of quantum encoding: quantum decoding. By interacting (non-destructively) with the encoded quantum state via auxiliary qubits we are able to determine the signature of errors that have affected the state, known as the syndrome. This process, syndrome extraction, offers incomplete information -- which itself may be unreliable. As such, we employ a decoding algorithm to determine the most likely error occurrences. Given the syndrome observed, the decoder outputs a best guess for the error that caused it, or alternatively a likely correction that will undo it. A family of error correcting codes may have a variety of decoding algorithms to choose from; picking the decoder is a balance between accuracy, speed, and compute budget for decoding.

These algorithms are purely classical and are typically deployed on hardware spatially non-local to the qubits. The speed of this process -- syndrome extraction and decoding -- defines the logical clock rate of the computer, and therefore the speed at which we can perform logical calculations \cite{skoric_parallel_2023, tan_scalable_2022}. Having touched upon the need for speed, let us now stress accuracy: a more accurate decoder will be more effective at producing a best guess for errors and corrections, and this will result in improved logical accuracy.  The decoder is a key element in the performance of the quantum error correction protocol. On top of these demands on speed and accuracy, a decoder is also bandwidth-thirsty.  Syndromes are extracted at the qubit level and communicated up the quantum stack --  any latency incurred during this communication subtracts from the time budget we have to decode the syndrome.

In this paper we propose a new two-stage decoding scheme. In the first stage, we employ belief propagation (BP) -- a widely used algorithm for decoding classical low-density parity-check codes (LDPC) \cite{geffner_reverend_2022}, that is not guaranteed to fully decode syndromes on quantum codes. In the cases where BP fails to fully decode a syndrome, we accept some corrections based on the algorithm output. We call this a partial correction, and update the syndrome accordingly. If there are remaining unexplained syndromes bits, these are passed to another decoder in a second stage to finalise the correction. In principle any conventional quantum decoder can be employed in this step; here we use minimum-weight perfect matching (MWPM).

Similar ideas for `pre-decoding' schemes were explored by Delfosse \cite{delfosse_hierarchical_2020} where a simple algorithm, based on local rules, attempts to explain an observed syndrome, calling a `back-up' decoder in the event of failure. Whenever a `fail' flag was shown however, all information from the local decoder was discarded. This idea was extended in Smith \emph{et. al} \cite{smith_local_2022}, with a particular focus on decoding cycle speed up.

In Ref.~\cite{meinerz_scalable_2022} a convolutional neural network (CNN) was used to partially correct errors that again were spatially local. Additional work exploring the use of CNNs in such pre-decoding schemes is contained in Refs.~\cite{ueno_neo-qec_2022, chamberland_techniques_2022}. 

Our scheme provides similar partial corrections to that of Refs.~\cite{meinerz_scalable_2022, smith_local_2022}, however we do not rely on local features or machine learning techniques, and instead make use of information across the entire QEC code.

BP has shown some promise on quantum codes.  However, due to its converge not being guaranteed, BP must always be paired with another decoder. In Ref.~\cite{panteleev_degenerate_2021, Roffe_2020} BP was used together with ordered statistics decoding to decode hypergraph product quantum LDPC codes. In Ref.~\cite{old_generalized_2022} BP was used as a partial decoder of itself, at a cost in accuracy compared to MWPM. In Ref.~\cite{criger_multi-path_2018, higgott_fragile_2022} BP was used to refine the input error model of conventional surface-code decoders, such as MWPM and Union-Find (UF), and an accuracy boost was reported compared to the standalone use of the conventional decoder. These decoders were called `belief-matching' and `belief-find'. In these schemes, as in ours, if BP converges then there is no need for the second stage. In the cases where BP does not converge, rather than simply informing the second stage we instead combine corrections achieved from both BP and the second stage decoder.

We benchmark our decoder on the rotated planar surface code under circuit-level noise, and find that BP is able to correct a significant portion of the observed syndrome. This reduces the computational load on the second stage decoder, and increases the overall accuracy compared to just using the conventional decoder. As we will explain, this boost in accuracy comes from the fact that the BP algorithm runs on a more accurate error model (one based on `decoding hypergraphs') rather than the one used by the conventional decoder  (one based on `decoding graphs'). When the conventional decoder is MWPM, the accuracy of our scheme sits half-way between the accuracy of MWPM and the higher accuracy of belief-matching. No previous partial decoder proposal has provided a timing analysis for circuit-level noise. Prior art either used syndrome weight reduction as a proxy for runtime improvement or only considered a toy phenomenological noise model.

In an architecture where BP is implemented in dedicated hardware close to the qubits, this reduced syndrome will require a smaller bandwidth to communicate to the second-stage decoder compared to what the original syndrome would require. Therefore, given such a close-to-qubit efficient implementation of BP, our scheme results in three types of gains compared to the single-stage conventional decoder: reduced bandwidth requirements, improved accuracy, and increased speed.

The remaining of this paper is organised as follows. In Section~\ref{sec:decoding_quantum_errors} we review the current state of the art surrounding decoders and, more specifically, those schemes relevant to partial decoding. We also clarify some of the terminology in use through this piece. In Section~\ref{beliefprop} we review the belief propagation algorithm and the minor adjustments required for it to provide solutions to a decoding problem. In Section~\ref{sec:BPasPD} we give a detailed description of our decoder. In Section~\ref{analysis} we showcase our numerical results on the rotated surface code.  We discuss our findings in Section~\ref{sec:conclusions}.

\section{Decoding and partial decoding}
In this section we will briefly review the standard approach to decoding quantum codes, with a particular focus on correlated errors. We will also review the belief propagation algorithm, before outlining our decoding scheme in more detail.

\subsection{Decoding quantum errors} \label{sec:decoding_quantum_errors}
In its simplest format, the decoding problem requires only two ingredients: a matrix $H$ and a vector $\mathbf{s}$. The parity check matrix $H$ has dimension $(n,m)$, where the rows define the $n$ stabiliser checks of the code, and the columns define $m$ error mechanisms. We set $H_{i, k}=1$ iff the stabilizer check $i$ anticommutes with error mechanism $k$, and $0$ otherwise. We say that this check is `triggered' by the error. The syndrome vector $\mathbf{s}$ is a binary vector such that $s_i=1$ if check $i$ has been triggered, and $0$ otherwise. Decoding reduces to determining a vector of error mechanisms  $\mathbf{e}$, such that $H \mathbf{e}=\mathbf{s}$.

Once we have our prediction of errors $\mathbf{e}$, rather than correcting each of them, we are instead interested only in their net effect. For the purposes of fault tolerant quantum computing, we can restrict ourselves to tracking the state of the logical observables we have encoded. We define a \emph{logical check} matrix $L$, where rows correspond to logical observables and columns correspond to error mechanisms, such that $L_{\alpha, k}=1$ iff observable $\alpha$ is flipped by error mechanism $k$. The state of the logical observables are  fully described by the homology vector $\pmb{\lambda}$ = $L \mathbf{e}$. This vector has elements $\lambda_\alpha=1$ iff logical $\alpha$ anticommutes with our predicted error pattern, and $0$ otherwise. In short, if our homology vector is the $0$ vector, no action must be taken. If the $\alpha$th element is $1$, we must adjust the logical Pauli frame for observable $\alpha$ in software.
The problem remains of how best to determine our prediction $\mathbf{e}$. Note that the check matrix $H$ is in general not invertible, and we must employ various decoding algorithms to make our prediction. 

\subsubsection*{Decoding graphs, hypergraphs, and correlated errors}

Quantum errors can be described as a random sequence sampled from the Pauli group $\{\mathds{1}, X, Y, Z\}$. With the fact that $Y = i XZ$, we therefore need only deal with $X$ and $Z$ errors in general. To date the most successful quantum codes are Calderbank-Shor-Steane (CSS) codes, such as the surface code, for which $X$-type and $Z$-type errors can be treated independently. Importantly for surface codes both $X$ and $Z$ error models are \emph{graphlike}, such that a single error anticommutes with at most two stabiliser checks. It is therefore instructive to employ \emph{decoding graphs} of the form $G=\{E, V\}$. 
Each edge $e\in E$ represents an error mechanism, and the set of vertices incident on that edge correspond to the stabiliser checks that anticommute with the error mechanism. The elements of our syndrome $\mathbf{s}$ are vertices on the decoding graph; to decode we must find a minimal set of edges that intersect all syndrome bits. Matching decoders such as MWPM operate on both $X$ and $Z$ decoding graphs simultaneously; however this practice does not use all information available.
In this setting, $Y$-type errors provide symptoms in both $X$ and $Z$ syndromes and therefore the decoding graphs are `correlated'. 
These errors contain information that should not be, but often is, discarded. Such an error mechanism can be referred to as a \emph{hyperedge}: an ``edge'' incident on more than two vertices. Our decoding graph then becomes a decoding \emph{hypergraph}.

Employing a decoding hypergraph precludes us from using matching decoders, however there are some decoders that have been developed to take this correlated information into account. 
The neural network decoder of Ref.~\cite{meinerz_scalable_2022} noted that the accuracy boost was due to avoiding this separation of errors into $X$ and $Z$ classes. For an intuitive picture on how decoding on hypergraphs can be performed, we require another graph-like abstraction.

\subsubsection*{Tanner graphs}

Tanner graphs are bipartite graphs $G_T=\{E, U\oplus V\}$ where the disjoint sets of vertices correspond to checks and error mechanisms. Edges connect error nodes to the check nodes that the error anticommutes with.
The biadjacency of this graph defines the parity check matrix $H$ and can represent complex error mechanisms in a graphlike fashion.
As each error mechanism is a vertex rather than an edge, hyperedges on decoding hypergraphs become error nodes connected to multiple check nodes.

In this picture, our syndrome vector is defined on the check nodes $U$. To decode we must find a minimal set of error nodes in $V$ that form a subgraph, or collection of subgraphs, with the syndrome. Up to an agreed ordering, this is equivalent to determining a vector $\mathbf{e}$ to satisfy $H\mathbf{e}=\mathbf{s}$. Tanner graphs (or factor graphs) are used extensively in classical decoding, and provide an intuitive picture to explore belief propagation: an efficient decoder with some application to quantum codes.

\subsection{Belief  propagation}\label{beliefprop}
Belief propagation is an efficient decoding algorithm for classical LDPC codes \cite{mackay_good_1999}. With a given input syndrome, BP can be understood as a heuristic implementation of Bayes' rule that updates the prior probabilities for statistically independent error mechanisms.

BP takes three inputs: a parity check matrix $H$, defining the Tanner graph of our error model, the syndrome $\mathbf{s}$, and a vector of prior probabilities for the error mechanisms, $\mathbf{p}$. These probabilities act as a `soft decision' vector: its elements represent the relative likelihood of each error occurring. The algorithm iteratively updates this decision vector by passing `beliefs' between check nodes and error nodes on the Tanner graph with information gained from the syndrome. At the end of each iteration, the soft decision vector is inspected and some errors are committed as a `hard decision'.
We commit all errors whose posterior probability odds are greater than or equal to 1 -- i.e., their probability of having occurred is larger than $0.5$. 
These error mechanisms form our predicted error vector $\mathbf{e}$, and if $H \mathbf{e} = \mathbf{s}$ then we say BP has \emph{converged}. Upon convergence, the algorithm terminates as we have explained our syndrome. If not, the next iteration of message passing begins and we continue until either convergence happens or a preset maximum number of iterations $m_{\mathrm{iter}}$ have occurred, which is a free parameter of the algorithm.

While BP is a highly successful decoder for classical codes, it is not guaranteed to converge on quantum codes due to quantum degeneracy. 
Syndrome patterns on quantum codes can be explained by different error mechanisms which are equivalent up to stabilisers, and at times BP can commit all of these as hard decisions \cite{poulin_iterative_2008}. The effect of this is that we commit to errors which explain syndrome bits, but equivalent errors are also committed, and we effectively reintroduce the explained syndrome bits. Without modifications there is no way for BP to escape such local-minima-type effects, and this is often referred to as a split-belief. As such, BP does not permit a threshold on quantum codes and to be useful must be paired with another decoder \cite{panteleev_degenerate_2021,old_generalized_2022, Roffe_2020, criger_multi-path_2018, higgott_fragile_2022}.

\subsubsection*{Algorithm details}
One iteration of BP consists of four stages: messages sent from checks to errors, messages sent from errors to checks, posterior probability calculation and a convergence check. Upon receipt of these messages, calculations are performed depending on the setting of the algorithm. Two of the most well known message updating schemes are min-sum and product-sum (sometimes referred to as tanh). Here, we employ product-sum updates as it has been shown to perform slightly better on quantum codes \cite{acharya_suppressing_2022}.

We will refer to error nodes as $e_k$ and check nodes as $c_i$. For convenience messages passed from check node $c_i$ to error node $e_k$ will be denoted $P_{i,k}$. Returning messages, from node $e_k$ to $c_i$, as $Q_{k, i}$. Messages are only passed between connected nodes, i.e., when discussing messages between $c_i$ and $e_k$ it is implicit that $H_{i, k}=1$. The syndrome bit for check node $c_i$ is denoted $s_i$; similarly the prior probability for an error $e_k$ occurring is $p_k$. It is useful to define \emph{node neighbourhoods} as the set of all error (check) nodes a given check (error) node is connected to. This is equivalent to isolating the non-zero indices in a specific row (column) of the parity check matrix $H$. The error neighbourhood of check node $c_i$ is denoted \[\mathcal{E}_i=\{e_k:H_{i, k}=1\},\] and the check neighbourhood of error node $e_k$:\[\mathcal{C}_k=\{c_i:H_{i, k}=1\}.\]

Upon initialisation, we set all $Q_{k, i}=\frac{p_k}{1-p_k}$. Each iteration undergoes the following steps:
\begin{enumerate}
    \item We pass all check-to-error messages:
    \begin{align}\label{eq:product_sum_check_to_error}
    \nonumber P_{i, k} &= \frac{1-\delta_{i,k}}{1+\delta_{i,k}},~\mathrm{where}\\
    \delta_{i,k} &= (-1)^{s_i}\prod_{ e_l\in\mathcal{E}_i\backslash\{e_k\}}\left(\frac{1-Q_{l, i}}{1+Q_{l, i}}\right).
    \end{align}
    Where here the set $\mathcal{E}_i\backslash\{e_k\}$ refers to all error nodes in the neighbourhood of $c_i$, except $e_k$. 
    \item Once all check-to-error messages have been updated, we pass all error-to-check messages:
    \begin{align}\label{eq:error_to_check}
 Q_{k, i} &= \frac{p_k}{1-p_k}\prod_{ c_j\in\mathcal{C}_k\backslash\{c_i\}}P_{j, k}.
    \end{align}
    \item As all messages have been passed between neighbouring nodes we now calculate the posterior probability odds for each error node $e_k$:
    \begin{equation}
    \chi_k = \frac{p_k}{1-p_k}\prod_{ c_i\in\mathcal{C}_k}P_{i,k},
    \end{equation} where the posterior odds relate to the posterior probability $p^\prime_k$ as $\chi_k\equiv \frac{p^\prime_k}{1-p^\prime_k}$. We then make a hard decision:
    \begin{equation}\mathbf{e}_k = \begin{cases}
        1 & \mathrm{if}~~\chi_k\geq 1,\\
        0&\mathrm{otherwise}.
    \end{cases}\end{equation}
    \item Finally, we check for convergence. If $H\mathbf{e}= \mathbf{s}$ we say we have converged and return $\mathbf{e}$ as the error pattern. If not, we start a new iteration and continue passing messages.
\end{enumerate}

As mentioned the program terminates only if we converge or a maximum number of iterations $m_{\textrm{iter}}$ have occurred. We implemented the `parallel' version of the BP algorithm -- in which all check-to-error messages are updated before error-to-check messages. An alternative construction, serial BP \cite{kuo_refined_2020}, iterates through the error nodes and for each node $e_{k^\prime}$ first updates all $P_{i, k=k^\prime}$, before all $Q_{k=k^\prime, i}$. In this work we only consider parallel updates.

The product-sum updates here correspond to Eq.~\eqref{eq:product_sum_check_to_error}. If we wish to modify the algorithm to use min-sum updates,  we simply need a different formulae for the check-to-error messages, see Ref.~\cite{roffe_decoding_2020}. The rest of the algorithm remains unchanged.

In the event BP does not converge, the final error pattern $\mathbf{e}$ does not explain the syndrome and is usually discarded. In other two stage decoding schemes the posterior probability vector $\mathbf{p}^\prime$ is used to inform the decoder that follows \cite{criger_multi-path_2018, higgott_fragile_2022}, however here we propose an alternative.

\subsection{Belief propagation as partial decoder}\label{sec:BPasPD}

\newcommand{\epartial}{\mathbf{e}^p}
\newcommand{\spartial}{\mathbf{s}^p}
\newcommand{\lpartial}{\pmb{\lambda}^p}
\newcommand{\econv}{\mathbf{e}^c}
\newcommand{\lconv}{\pmb{\lambda}^c}

We call a decoder \emph{partial} if its output is not guaranteed to fully explain the input syndrome, and it instead computes a \emph{partial} correction -- errors that the decoder is confident have occurred. To find a full solution, the partial decoder can be combined with a conventional decoder.

Having specified the parity check matrix $H$ and the error probabilities $\mathbf{p}$, the input to a partial decoder is the observed syndrome $\mathbf{s}$. A partial decoder computes an error vector $\epartial$, which we refer to as a partial correction. From this we can compute a partial homology $\lpartial = L \epartial$, and an updated syndrome $\spartial = \mathbf{s} \oplus  \left( H \epartial \right)$,  where $\oplus$ denotes \verb"xor" or sum modulo $2$. Note $\spartial$ can be the zero vector when the partial decoder succeeds in decoding the syndrome fully. Recall that for fault-tolerant computation we only need to track logical errors, it is therefore sufficient for the partial decoder to output the partial homology and updated syndrome. If the updated syndrome is non-empty, the conventional decoder computes a correction $\econv$ such that $H \econv=\spartial$, and the corresponding homology result $\lconv$. The final homology vector  is therefore $\lpartial \oplus \lconv$.

We propose using BP once as a partial decoder and, when BP does not converge, combining it with a conventional decoder. In particular, here we explore BP combined with MWPM (BP+MWPM), which is given in Algorithm \ref{alg:pseudocode}. In addition to the BP attribute $m_{\textrm{iter}}$, we define a tolerance $t_{\textrm{BP}}$ attribute. Given the observed syndrome and initial error probabilities $\mathbf{p}$, BP computes a vector of updated error probabilities $\mathbf{p}^\prime$, where $p_k^\prime$ is the posterior probability for the $k$th error in the error vector $\mathbf{e}$. BP as a partial decoder then computes the partial correction $\epartial$ as 
\begin{equation*}
    \mathbf{e}^p_k = 
    \begin{cases}
        1, & \text{if}\ p_k^\prime \geq t_{\textrm{BP}} \\
        0, & \text{otherwise}
    \end{cases}
    .
\end{equation*}

\begin{algorithm}[h]
    \caption{Pseudocode for BP+MWPM. BP is assumed to use the product-sum algorithm.} 
    \label{alg:pseudocode}
    \begin{algorithmic}[1]
        \Function{BP+MWPM}{$\mathbf{s}, H, \mathbf{p}, L, m_{\textrm{iter}},  t_{\textrm{BP}} $}
        \item[]
        \LeftComment{BP computes the updated error probabilities}
        \State $\mathbf{p}^\prime \gets \textrm{BP}\left(\mathbf{s}, H, \mathbf{p}, m_{\textrm{iter}} \right)$
        \item[]
        \LeftComment{Compute the partial correction $\epartial$}
        \For{$k \gets 0$ to size of $\mathbf{p}^\prime$}
            \If{$p^\prime_k \geq t_{\textrm{BP}} $}
                \State $e_k \gets 1$
            \Else
                \State $e_k \gets 0$
            \EndIf
        \EndFor
        \item[]
        \LeftComment{Compute the updated syndrome}
        \State $\spartial \gets \mathbf{s} \oplus \left( H  \epartial \right)$
        \LeftComment{Compute the partial homology}
        \State $\lpartial \gets L  \epartial$
        \item[]
        \If{$\spartial$ is empty}
            \State \textbf{return} $\lpartial$
        \Else
            \State $\econv \gets \textrm{MWPM}\left(\spartial, H, \mathbf{p} \right)$
            \State $\lconv = L  \econv$
            \State \textbf{return} $\lpartial \oplus \lconv$
        \EndIf
        \EndFunction
    \end{algorithmic}
\end{algorithm}

A motivation for using BP as a partial decoder is that BP can decode over a parity check matrix representing a decoding hypergraph. In contrast, decoding algorithms such as MWPM and UF require the parity check matrix to represent a decoding graph -- with edges with degree at most $2$. As discussed in Section \ref{sec:decoding_quantum_errors}, this means that BP can directly detect independent errors that in the conventional decoding graph might only be represented as combinations of other independent errors, which suppresses the representation of their probability. Making use of the more refined error model means that BP as a partial decoder combined with a conventional decoder boosts the accuracy of the conventional decoder.

As discussed above, decoders also need to be fast. BP is a well known classical algorithm with implementations optimised for speed \cite{nikishkin_bp_wireless}. In the cases when BP does not converge, BP as a partial decoder outputs an updated syndrome with a significantly smaller Hamming weight than the original syndrome. Recall that the Hamming weight is the number of non-zero elements in a string -- in our context, the number of non-trivial elements in the syndrome. For decoding algorithms such as MWPM the speed depends on the size of the syndrome \cite{higgott_pymatching_2021, fowler_minimum_2014}. Therefore, the execution time of the conventional decoder is reduced when BP is used as a partial decoder. In the cases when BP converges, the conventional decoder does not need to be executed at all.

Given the simplicity and the message-passing nature of the BP algorithm, we envisage an implementation of BP in hardware close to the qubits. Reducing the Hamming weight of the syndrome passed to the conventional decoder would lower the bitrate requirements. The partial homology bit vector scales in length with the number of logical observables, therefore passing this information through the control stack should also be cheap.

\section{Analysis}\label{analysis}

Here we discuss the results of our simulations. We simulate a logical Pauli-$Z$ memory on the rotated surface code under circuit-level noise. We employ the product-sum version of BP, setting $m_{\textrm{iter}}=30$, and $t_{\textrm{BP}}=0.9$. For more details see Section~\ref{sec:implementation}. 

\subsection{Simulation results} \label{sec:sim_results}

\begin{figure}[t]
    \centering
    \includegraphics[width=220pt]{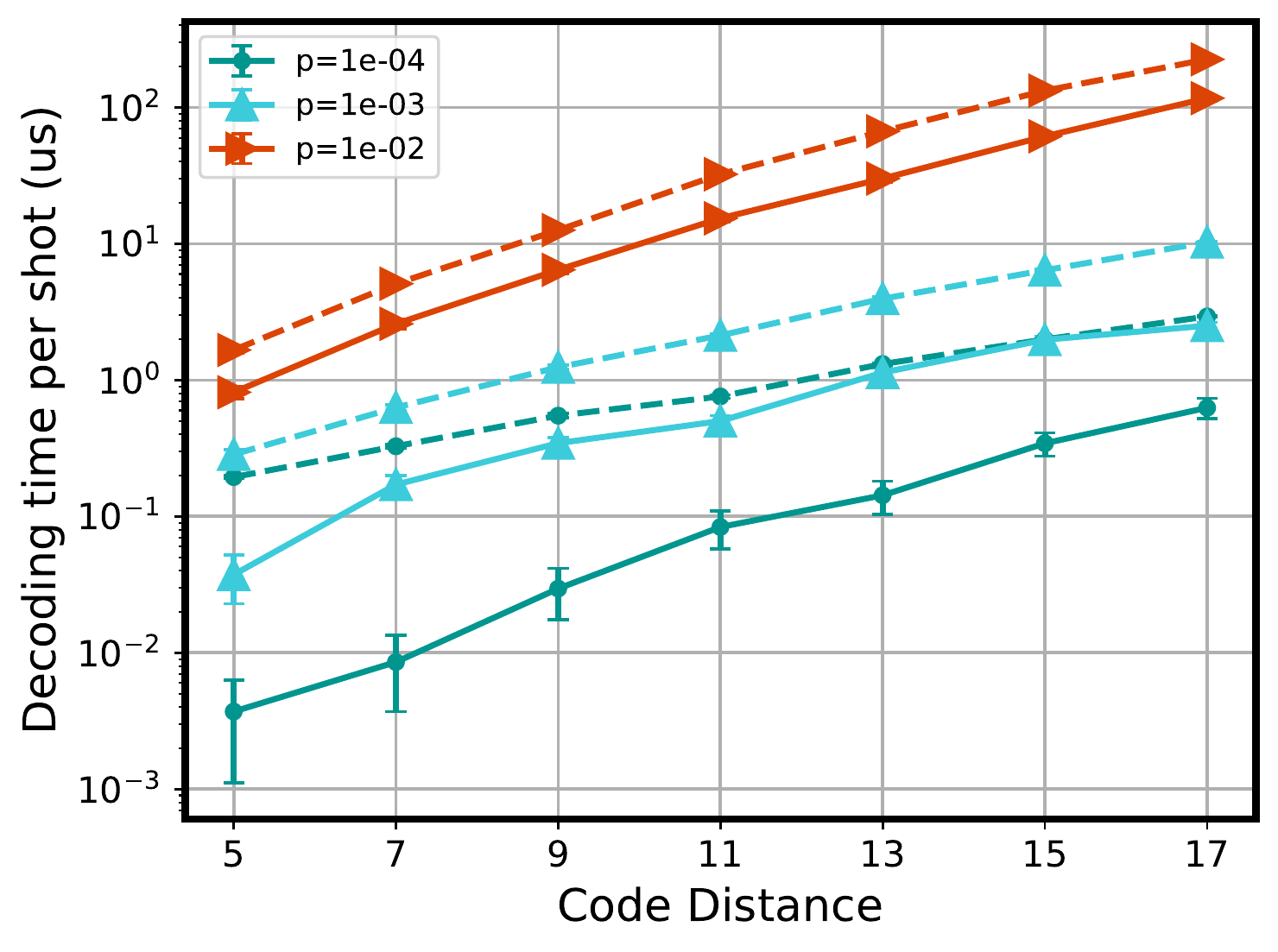}
    \caption{Mean MWPM decoding time per shot with partial decoding (solid lines) and without (dashed lines), for physical error rates $1\%$ (red arrowheads), $0.1\%$ (blue triangles) and $0.01\%$ (green dots) and a range of distances. BP as partial decoder parameters are $m_{\textrm{iter}}=30$ and $t_{\textrm{BP}}=0.9$.}
    \label{fig:speedup}
\end{figure} 

Our discussion assumes an implementation in which BP runs on fast, dedicated hardware physically close to the qubits; whereas MWPM runs on a separate component further up the quantum computational stack. Throughout we will compare the runtime of MWPM in schemes both with and without partial decoding.

In other works the size of the syndrome that is passed to the conventional decoder is used as a proxy for its runtime \cite{chamberland_techniques_2022}. Here we explicitly time the MWPM decoder using the techniques outlined in Ref.~\cite{higgott_sparse_2023}, with more details in Sec.~\ref{sec:implementation}.

\subsubsection*{Speeding up the decoding cycle}
In Fig.~\ref{fig:speedup} we plot the mean time taken for MWPM to decode a syndrome from a surface code of distance $d$ and physical error rate $p$. The solid lines indicate the results using partial decoding, whereas the dashed lines relate to a standard implementation of MWPM.
For a high physical error rate of $p=1\%$ our scheme exhibits an approximately constant speed up of $2$x. In this regime syndromes are very dense and BP is throttled by split beliefs. As we decrease the physical error rate to $p=0.1\%$ however, we begin to see the impact of the code distance. For $d=5$ our scheme speeds up MWPM by over $7$x, and as the distance increases this value stabilises to approximately $3.75$x. In a regime where the syndromes are very sparse, $p=0.01\%$, we observe a dramatic speedup for low distances -- $52$x faster than MWPM for $d=5$. This difference decreases with increasing distance and converges towards $4$x. Notably, when using our scheme with physical error rate $p=0.1\%$ we are consistently faster, or as fast, as using standalone MWPM on experiments with physical error rate $p=0.01\%$.

\begin{figure}
    \centering
    \includegraphics[width=220pt]{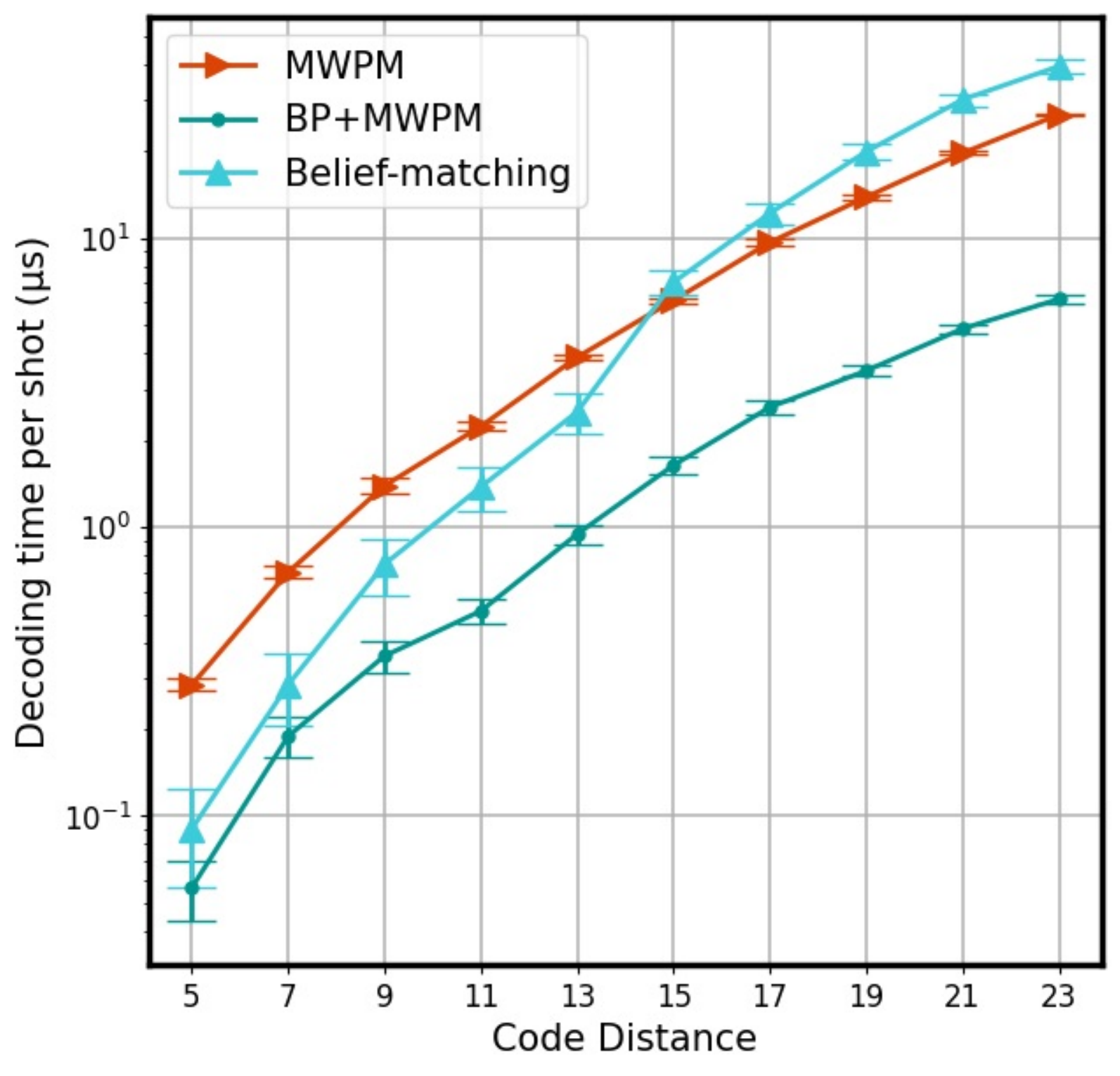}
    \caption{Mean decoding time per shot of the MWPM step of the decoder for three decoding algorithms: MWPM, belief-matching and BP+MWPM with circuit-level noise and $p_{\textrm{phys}}=0.1\%$. BP parameters are $m_{\textrm{iter}}=30$ and $t_{\textrm{BP}}=0.9$.}
    \label{fig:speed_comparison}
\end{figure}

In Fig.~\ref{fig:speed_comparison} we fix the physical error rate to $p=0.1\%$ and compare the mean decoding time of three different schemes, each of which utilise MWPM. As a baseline we plot MWPM as a standalone decoder (red line). We then plot the MWPM portion of belief-matching as described in \cite{higgott_fragile_2022} (blue line), and the MWPM stage of our scheme (green line). One can see that, while our scheme is speeding up MWPM by roughly $3.75$x, belief-matching begins to become slower compared to MWPM as distance increases. While resulting in more accurate matchings, suprisingly, the new edge weights from the BP posterior probabilities slow down the MWPM stage of belief-matching.  This increased runtime is on top of any latency incurred in while updating all edge weights at every shot -- a latency that BP as partial decoder avoids, as it does not update the error model of the conventional decoder at every shot.

\subsubsection*{Syndrome reduction}
\begin{figure}[t]
    \centering
    \includegraphics[width=220pt]{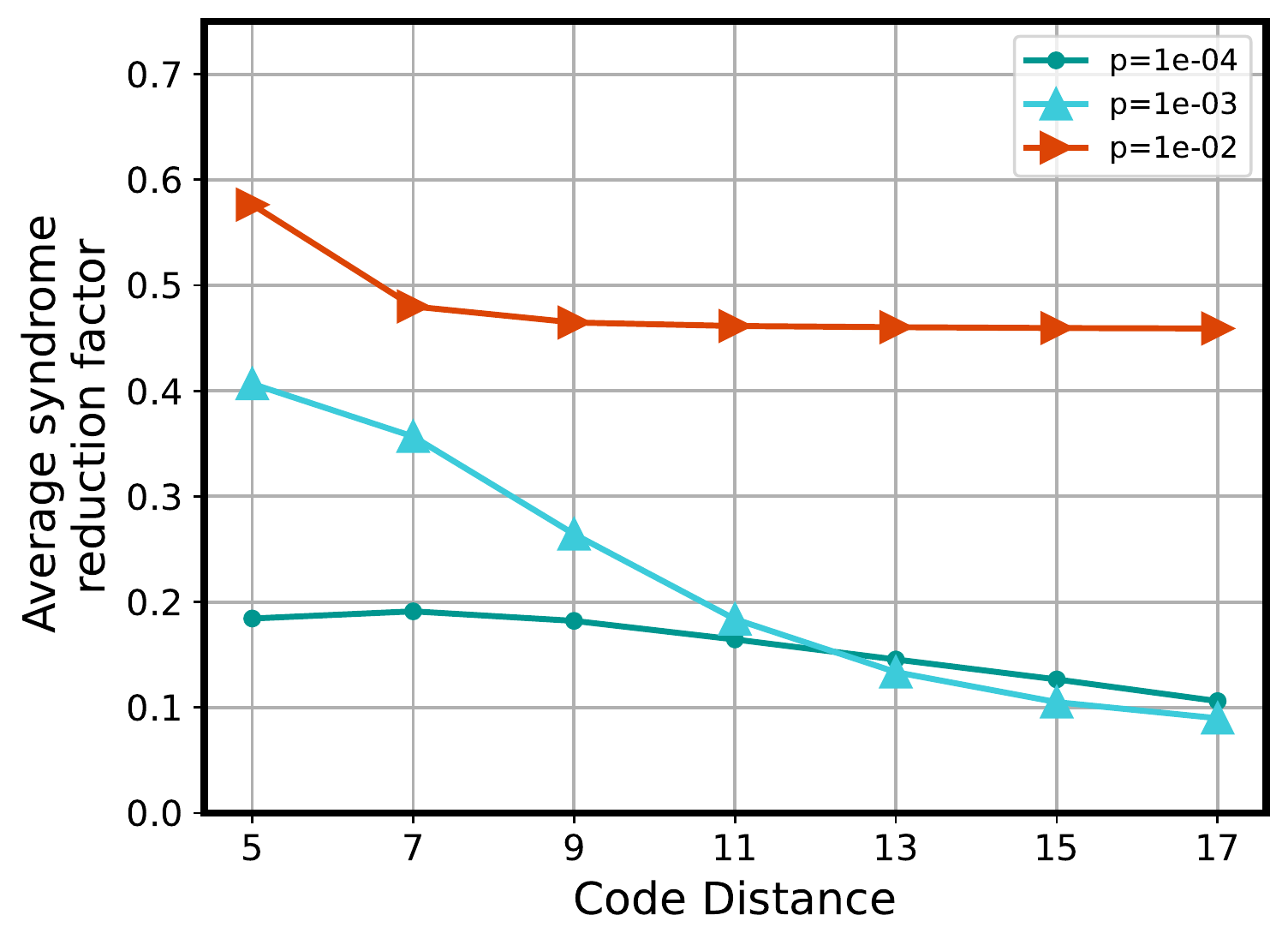}
    \caption{Average syndrome reduction, here calculated as the ratio between the syndrome Hamming weight with and without partial decoding. Plotted over a range of distances, the lines correspond to physical error rates $1\%$ (red arrowheads), $0.1\%$ (blue triangles) and $0.01\%$ (green dots). BP as partial decoder parameters are $m_{\textrm{iter}}=30$ and $t_{\textrm{BP}}=0.9$.}
    \label{fig:bandwidth}
\end{figure}

To better understand this speedup, and how it relates to the syndromes being passed to MWPM, in Fig.~\ref{fig:bandwidth} we highlight what proportion of the syndrome we are correcting with BP. That is, we plot the ratio of the syndrome Hamming weight after partial decoding and prior. This is a direct analogue to how many bits of information must be passed further up the quantum stack. We can see that for large physical error rates the effect is muted. While we are able to reduce the syndrome to less than a half of its original length, the high noise capacity is not amenable for the BP algorithm. Moving to noise levels far below threshold we see pronounced effects of using partial decoding. For $p=0.1\%$ we see significant improvement as distance increases, where for high distances it seems to converge to a point around 0.1 -- meaning we have reduced the number of bits we are required to send by $90\%$. For very low physical error rates $p=0.01\%$ we are always reducing the syndrome to less than one fifth of its original weight.

\begin{figure}[t]
    \centering
    \includegraphics[width=220pt]{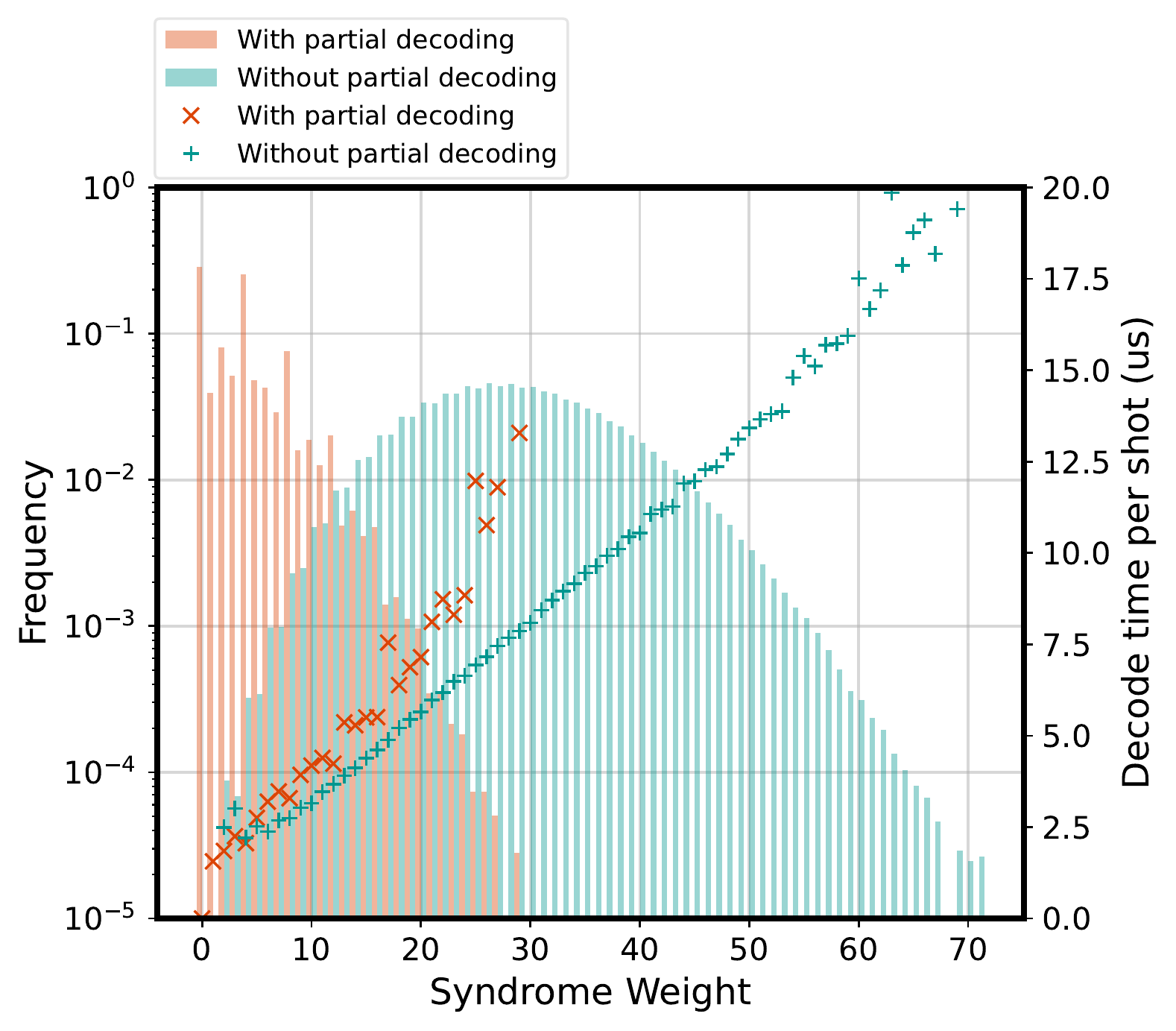}
    \caption{{\bf Left $y$ axis:} Log-scale histograms of syndrome Hamming weights for $d$ rounds of a distance $d=13$ quantum memory experiment, under $p_{\textrm{phys}}=0.1\%$ noise. The red distribution represents the distribution of syndrome weights passed to MWPM in the partial decoding scheme. \newline
    {\bf Right $y$ axis:} The runtime of MWPM as a function of the syndrome weight without (green) and with (red) partial decoding.}
    \label{fig:raw_data}
\end{figure}

  For an insight into the raw data we collected during these simulations, Fig.~\ref{fig:raw_data} shows two plots overlayed for $d=13$ and $p=0.1\%$. On the left y-axis we show a log-scale histogram of syndrome weights being communicated to MWPM both without partial decoding (green bars) and with partial decoding (red bars). The shift in the syndrome distribution is significant, and, to our resolution, the zero weight bar is populated only for the partial decoding scheme. In Appendix~\ref{app:histograms} we display more histograms for a broader range of experiment parameters.

Next, we ask to what extent syndrome reduction can be interpreted as a good proxy for decoding speed up. On the right y-axis of Fig.~\ref{fig:raw_data} we are plotting the decoding time per shot in MWPM as a function of  syndrome weight, both with (red markers) and without (green markers) partial decoding. Notably the time taken to decode a syndrome of weight $w$ is slightly longer if it has been partially decoded. This is related to the shift in syndrome distributions: BP reduces a syndrome of weight $w_1$ to one of $w_0$, which does not necessarily coincide with an error pattern that would have naturally conferred a syndrome of weight $w_0$. As such, MWPM has a more difficult job to decode what may be an unusual syndrome pattern. As we will see however, this penalty does not negatively affect our logical accuracy. Since syndrome weight alone does not determine decoder running time, we conclude syndrome reduction should not be used as a precise proxy metric for speed-up. Nevertheless, syndrome reduction is the stronger effect and overall a speed-up is observed (recall Fig.~\ref{fig:speedup} and Fig.~\ref{fig:speed_comparison}). 

\subsubsection*{Improved accuracy}

\begin{figure}
    \centering
    \includegraphics[width=220pt]{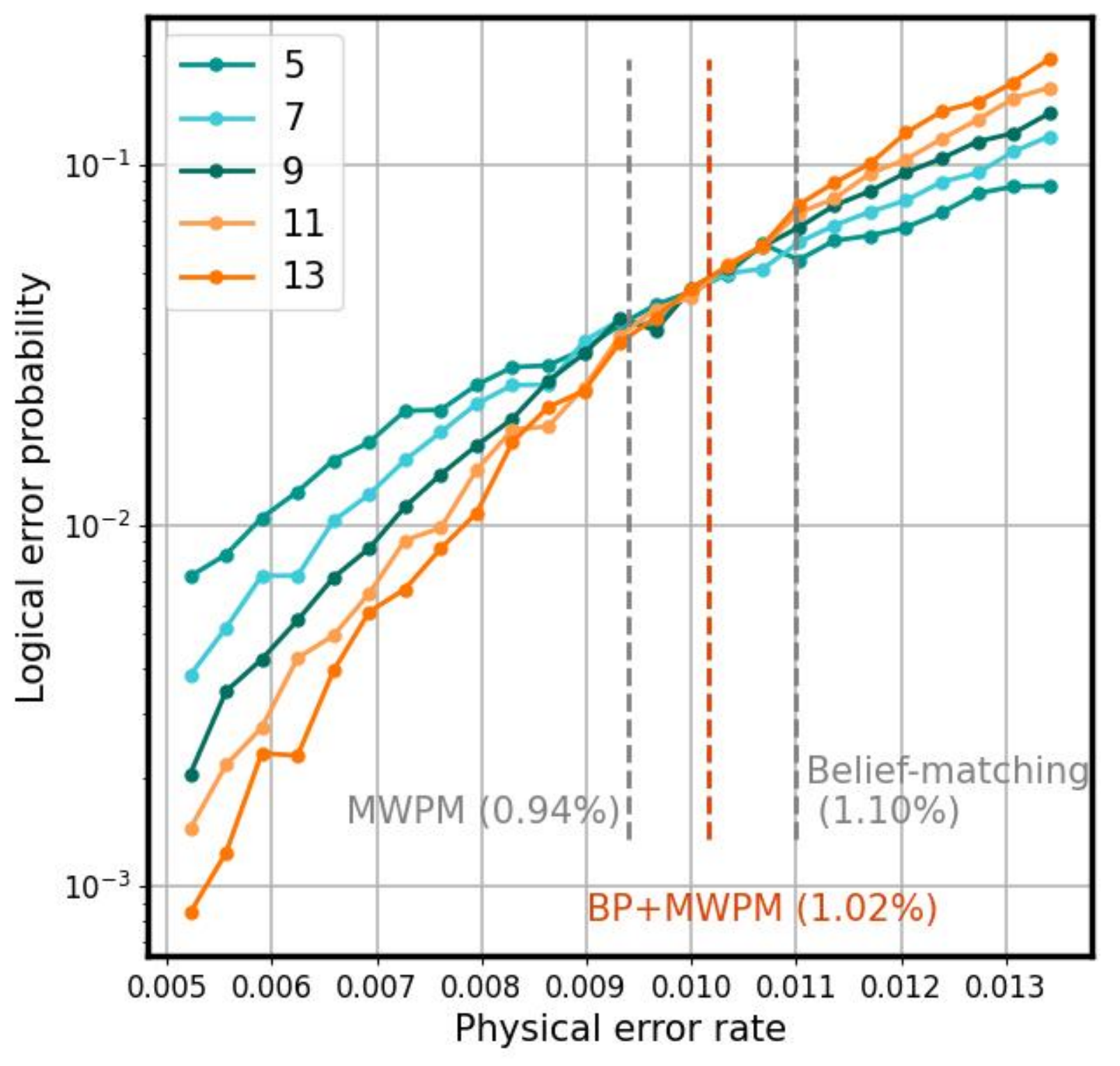}
    \caption{Threshold plot of the surface code $Z$-memory decoded with BP+MWPM, $t_{\textrm{BP}}=0.9$ and $m_{\textrm{iter}}=30$ under circuit-level noise. For reference we include the thresholds of MWPM and belief-matching. For their threshold plots see Appendix \ref{app:threshold_plots}.}
    \label{fig:accuracy}
\end{figure}

Finally, we examine the accuracy of our decoder in Fig.~\ref{fig:accuracy}, and compare it with MWPM and belief-matching.  We observe that adding the BP partial decoder to MWPM raises the threshold error rate from $0.94$\% to $1.02$\%, while belief-matching has threshold error rate $1.10$\%. For comparison we show the threshold plots for MWPM and belief-matching in Appendix \ref{app:threshold_plots}.

\subsection{Implementation details}\label{sec:implementation}
We compare decoders based on two key metrics: logical accuracy of $Z$-memory experiment and runtime of MWPM, the common feature to the decoders evaluated here. The logical $Z$-memory experiment is simulated using native $Z$-basis resets and measurement and Clifford gates $\{H, CZ\}$. At each distance $d$, the number of measurement rounds is equal to $d$. We simulate the experiments with circuit-level noise. Specifically, given a physical error rate $p$, we apply:
\begin{itemize}
    \item single-qubit depolarising channel with probability $p/10$ on single qubit gates,
    \item two-qubit depolarising channel with probability $p$ on two qubit gates,
    \item single-qubit depolarising channel with probability $p/10$ on reset and measurement collapse operations,
    \item  flip each measurement result with probability $p$,
    \item single-qubit depolarising channel with probability $p/10$ on idle qubits.
\end{itemize}
In general, we repeat each experiment for an average of $\min\!{\left(10^6, 1/p^2\right)}$ shots.
\begin{figure*}[t!]
    \centering
    \includegraphics[width=440pt]{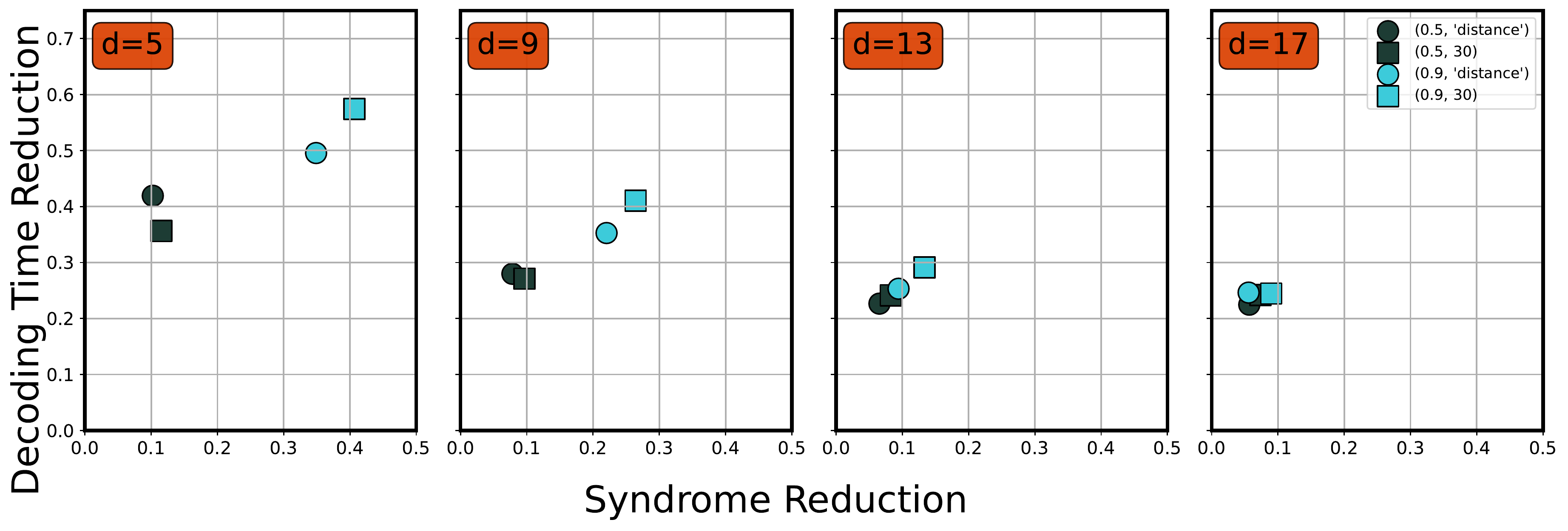}
    \caption{Average syndrome reduction and decoding speed-up for different code distances and $m_{\textrm{iter}}$ and $t_{\textrm{BP}}$ parameters.}
    \label{fig:hyperparams_converge}
\end{figure*}
We implement our BP+MWPM decoder, the belief-matching decoder, and standalone MWPM in Python with open source software. Specifically, we use the MWPM implementation in \textit{PyMatching 2}\footnote{\url{https://github.com/oscarhiggott/PyMatching}}\cite{higgott_sparse_2023} and the BP implementation in BP+OSD\footnote{\url{https://github.com/quantumgizmos/bp_osd}}\cite{Roffe_2020}.

As suggested in \cite{higgott_sparse_2023} we time the decoders in batches of $10^4$ syndromes. The MWPM step of belief-matching requires a new decoder object for each syndrome, since the weights of the decoding graph are updated after the BP step. This means we cannot decode $10^4$ different syndromes in a batch. Instead we sample 100 syndromes and average the runtime of decoding a batch of $10^4$ identical copies of each syndrome. We find that such approximation is comparable to the timings of MWPM obtained simply decoding a batch of $10^4$ different syndromes for the \textit{PyMatching 2} decoder. To avoid timing the process of storing the C\texttt{++} cache of the decoder we first decode once. Runtime is computed using a single core of an Apple M1 processor. We expect that the relative speed-ups reported here would hold up to dedicated hardware implementations of these decoders -- as long as the assumption of a negligible BP runtime applies, and barring any extra latencies incurred when updating the weights of the belief-matching decoder.

In the results we present here we have been using the parallel product-sum implementation of BP. As discussed in Section \ref{sec:BPasPD} the additional parameters of BP+MWPM are the number of iterations performed in the BP algorithm  $m_{\textrm{iter}}$ and the tolerance value $t_{\textrm{BP}}$. We ran and compared experiments with combinations of settings $m_{\textrm{iter}} \in \{d, 30, d^2 \}$ and $t_{\textrm{BP}} \in \{0.5, 0.9 \}$, where $d$ denotes the distance of a rotated planar code. Although the parameter $m_{\textrm{iter}}$ should scale with $d$, we included $m_{\textrm{iter}}=30$ following \cite{higgott_fragile_2022}. For the range of distances we explored (up to $d=13$), this represents a scaling of $m_{\textrm{iter}}$ between $d$ and $d^2$. Figure \ref{fig:hyperparams_converge} illustrates that as the code distance is increased the average decoding time reduction and average syndrome reduction converges for combinations of parameters $m_{\textrm{iter}} \in \{d, 30\}$ and $t_{\textrm{BP}} \in \{0.5, 0.9 \}$. We obtain higher threshold values when setting $t_{\textrm{BP}}=0.9$ compared to $t_{\textrm{BP}}=0.5$. We compute the highest threshold at values $t_{\textrm{BP}}=0.9$ and $m_{\textrm{iter}}=30$. We also note that depending on the noise in the device, choice of code and distance, the optimal values of $m_{\textrm{iter}}$ and $t_{\textrm{BP}}$ for accuracy could vary. Table \ref{tab:thrshld_values_hyperparams} summarises the threshold values computed with various $m_{\textrm{iter}}$ and $t_{\textrm{BP}}$ values.

\begin{table}[h]
\begin{center}
\begin{tabular}{|l||*{5}{c|}}\hline
\backslashbox{$m_{\textrm{iter}}$}{$t_{\textrm{BP}}$}
&\makebox[3em]{$0.5$}&\makebox[3em]{$0.9$}\\\hline\hline
$d$ & $0.96$ & $1.00$\\\hline
$30$ & $0.98$ & $\mathbf{1.02}$\\\hline
$d^2$ & $0.95$ & $0.98$\\\hline
\end{tabular}
\caption{Threshold values of surface code $Z$-memory under circuit-level noise decoded using BP+MWPM with various $m_{\textrm{iter}}$ and $t_{\textrm{BP}}$ values.}
\label{tab:thrshld_values_hyperparams}
\end{center}
\end{table}

\section{Discussion}\label{sec:conclusions}
We have presented a new two-stage quantum decoder.  In the first stage, a partial decoder based on BP is used to perform a partial correction of the error.  In the second stage, a conventional decoder is used to correct any remaining errors. We have benchmarked our decoder on the rotated surface code, using MWPM in the second stage.  Assuming a fast implementation of BP, our decoder offers a speed-up between $2$x-$4$x on average depending on the parameter regime and raises the threshold from $0.94\%$ to $v\%$ compared to a standalone use of the MWPM decoder.

One possible implementation of the BP partial decoder is in very close proximity to the qubits. An example of very fast light-weight logic co-local to qubits in superconducting systems is SFQ logic \cite{mcdermott_quantum--classical_2017}. Given the simplicity of BP, we expect it can be implemented under stringent compute budgets. 
Partial decoders on SFQ logic have been proposed before \cite{ueno_neo-qec_2022, ravi_better_2022}.

The input of the BP partial decoder is a digitised list of syndrome measurements (stabilisers that have detected errors or not) -- therefore, in the co-local implementation, we would assume that such digitisation is done at the QPU level.  This is currently not the standard in superconducting platforms, where analog measurement signals are sent higher up in the stack, where they are digitised \cite{krantz_quantum_2019, blais_circuit_2021}. There are, however, proposals and demonstrations for in-fridge measurement digitisation \cite{govia2014high, opremcak2021high, di2023discriminating}.

The output of the BP partial decoder is a reduced syndrome that needs to be communicated to the conventional decoder. Since our partial decoder reduces the density of errors, the BP partial decoder lowers bandwidth requirements between the first and second decoding stage, compared to the bandwidth that would be required to only use the conventional decoder (see histograms in Fig.~\ref{fig:app_histograms}). In this comparison we are assuming that the communication protocol between the two decoding stages only sends the addresses of the non-trivial syndromes that remain, as in \cite{smith_local_2022, ravi_better_2022}. Contrast this with a protocol that would communicate the measurement outcomes of all stabilisers -- if not compressed, this protocol would have to communicate an identical-length message no matter the sparsity of the remaining syndrome (a sparse syndrome would be a list with many zeros and just a few ones for the measurements that report errors) \cite{delfosse_hierarchical_2020, das_afs_2022}.

Communication across cryogenic stages of a dilution refrigerator is expensive, and accordingly there are several proposals in the literature for reducing the bandwidth required to communicate errors to the decoder \cite{delfosse_hierarchical_2020, das_afs_2022, smith_local_2022, ravi_better_2022}. We note, however, that the message we need to communicate to the conventional decoder is a digital message -- a syndrome.  While analog signals require amplification across cryogenic stages, which subtract from in-fridge power budget and add latency to the communication, there exist comparatively efficient solutions for cryogenic digital communication \cite{mcdermott_quantum--classical_2017}.

In our analysis we have neglected the time cost of BP, because we only expect this to be small when we have a highly parallelised implementation (e.g. using dedicated hardware).  However, valuable future work would be to build such a parallelised implementation and reassess our benchmarks. 

\subsection*{Acknowledgements}
We thank Ben Barber for illuminating discussions and insights, and Oscar Higgott for advice regarding implementing belief-matching.

We thank Joseph Rahamim, Matthew Hutchings, and Amir Salim from Seeqc Inc.~for insights about measurement digitisation.

This work was in part funded by project 10005792 from Innovate UK.

\bibliography{references.bib}

\begin{thebibliography}{10}

\bibitem{Dennis_2002}
Eric Dennis, Alexei Kitaev, Andrew Landahl, and John Preskill.
\newblock ``Topological quantum memory''.
\newblock \href{https://dx.doi.org/10.1063/1.1499754}{Journal of Mathematical
  Physics {\bf 43}, 4452--4505}~(2002).

\bibitem{campbell2017roads}
Earl~T Campbell, Barbara~M Terhal, and Christophe Vuillot.
\newblock ``Roads towards fault-tolerant universal quantum computation''.
\newblock \href{https://dx.doi.org/https://doi.org/10.1038/nature23460}{Nature
  {\bf 549}, 172--179}~(2017).

\bibitem{Roffe_2019_QEC}
Joschka Roffe.
\newblock ``Quantum error correction: an introductory guide''.
\newblock \href{https://dx.doi.org/10.1080/00107514.2019.1667078}{Contemporary
  Physics {\bf 60}, 226--245}~(2019).

\bibitem{skoric_parallel_2023}
Luka Skoric, Dan~E. Browne, Kenton~M. Barnes, Neil~I. Gillespie, and Earl~T.
  Campbell.
\newblock ``Parallel window decoding enables scalable fault tolerant quantum
  computation''~(2023).
\newblock arXiv:2209.08552 [quant-ph].

\bibitem{tan_scalable_2022}
Xinyu Tan, Fang Zhang, Rui Chao, Yaoyun Shi, and Jianxin Chen.
\newblock ``Scalable surface code decoders with parallelization in
  time''~(2022).
\newblock arXiv:2209.09219 [quant-ph].

\bibitem{geffner_reverend_2022}
Judea Pearl.
\newblock ``Reverend {Bayes} on {Inference} {Engines}: {A} {Distributed}
  {Hierarchical} {Approach}''.
\newblock In Hector Geffner, Rina Dechter, and Joseph~Y. Halpern, editors,
  Probabilistic and {Causal} {Inference}.
\newblock \href{https://dx.doi.org/10.1145/3501714.3501727}{Pages 129--138}.
\newblock ACM, New York, NY, USA~(2022).
\newblock 1 edition.

\bibitem{delfosse_hierarchical_2020}
Nicolas Delfosse.
\newblock ``Hierarchical decoding to reduce hardware requirements for quantum
  computing''~(2020).
\newblock arXiv:2001.11427 [quant-ph].

\bibitem{smith_local_2022}
Samuel~C. Smith, Benjamin~J. Brown, and Stephen~D. Bartlett.
\newblock ``A local pre-decoder to reduce the bandwidth and latency of quantum
  error correction''~(2022).
\newblock arXiv:2208.04660 [cond-mat, physics:quant-ph].

\bibitem{meinerz_scalable_2022}
Kai Meinerz, Chae-Yeun Park, and Simon Trebst.
\newblock ``Scalable {Neural} {Decoder} for {Topological} {Surface} {Codes}''.
\newblock \href{https://dx.doi.org/10.1103/PhysRevLett.128.080505}{Phys. Rev.
  Lett. {\bf 128}, 080505}~(2022).

\bibitem{ueno_neo-qec_2022}
Yosuke Ueno, Masaaki Kondo, Masamitsu Tanaka, Yasunari Suzuki, and Yutaka
  Tabuchi.
\newblock ``{NEO}-{QEC}: {Neural} {Network} {Enhanced} {Online}
  {Superconducting} {Decoder} for {Surface} {Codes}''~(2022).
\newblock arXiv:2208.05758 [quant-ph].

\bibitem{chamberland_techniques_2022}
Christopher Chamberland, Luis Goncalves, Prasahnt Sivarajah, Eric Peterson, and
  Sebastian Grimberg.
\newblock ``Techniques for combining fast local decoders with global decoders
  under circuit-level noise''~(2022).
\newblock arXiv:2208.01178 [quant-ph].

\bibitem{panteleev_degenerate_2021}
Pavel Panteleev and Gleb Kalachev.
\newblock ``Degenerate {Quantum} {LDPC} {Codes} {With} {Good} {Finite} {Length}
  {Performance}''.
\newblock \href{https://dx.doi.org/10.22331/q-2021-11-22-585}{Quantum {\bf 5},
  585}~(2021).

\bibitem{Roffe_2020}
Joschka Roffe, David~R. White, Simon Burton, and Earl Campbell.
\newblock ``Decoding across the quantum low-density parity-check code
  landscape''.
\newblock \href{https://dx.doi.org/10.1103/physrevresearch.2.043423}{Physical
  Review Research{\bf 2}}~(2020).

\bibitem{old_generalized_2022}
Josias Old and Manuel Rispler.
\newblock ``Generalized {Belief} {Propagation} {Algorithms} for {Decoding} of
  {Surface} {Codes}''~(2022).
\newblock arXiv:2212.03214 [quant-ph].

\bibitem{criger_multi-path_2018}
Ben Criger and Imran Ashraf.
\newblock ``Multi-path {Summation} for {Decoding} {2D} {Topological} {Codes}''.
\newblock \href{https://dx.doi.org/10.22331/q-2018-10-19-102}{Quantum {\bf 2},
  102}~(2018).

\bibitem{higgott_fragile_2022}
Oscar Higgott, Thomas~C. Bohdanowicz, Aleksander Kubica, Steven~T. Flammia, and
  Earl~T. Campbell.
\newblock ``Fragile boundaries of tailored surface codes and improved decoding
  of circuit-level noise''~(2022).
\newblock arXiv:2203.04948 [quant-ph].

\bibitem{mackay_good_1999}
D.J.C. MacKay.
\newblock ``Good error-correcting codes based on very sparse matrices''.
\newblock \href{https://dx.doi.org/10.1109/18.748992}{IEEE Transactions on
  Information Theory {\bf 45}, 399--431}~(1999).

\bibitem{poulin_iterative_2008}
David Poulin and Yeojin Chung.
\newblock ``On the iterative decoding of sparse quantum codes''~(2008).
\newblock arXiv:0801.1241 [quant-ph].

\bibitem{acharya_suppressing_2022}
Google~Quantum AI.
\newblock ``Suppressing quantum errors by scaling a surface code logical
  qubit''~(2022).
\newblock arXiv:2207.06431 [quant-ph].

\bibitem{kuo_refined_2020}
Kao-Yueh Kuo and Ching-Yi Lai.
\newblock ``Refined {Belief} {Propagation} {Decoding} of {Sparse}-{Graph}
  {Quantum} {Codes}''~(2020).
\newblock  url:~\url{http://arxiv.org/abs/2002.06502}.

\bibitem{roffe_decoding_2020}
Joschka Roffe, David~R. White, Simon Burton, and Earl~T. Campbell.
\newblock ``Decoding {Across} the {Quantum} {LDPC} {Code} {Landscape}''.
\newblock \href{https://dx.doi.org/10.1103/PhysRevResearch.2.043423}{Physical
  Review Research {\bf 2}, 043423}~(2020).

\bibitem{nikishkin_bp_wireless}
Pavel Nikishkin, Ruslan Goriushkin, Nikita Vinogradov, Evgeny Likhobabin, and
  Vladimir Vityazev.
\newblock ``High throughput fpga implementation of min-sum ldpc decoder
  architecture for wireless communication standards''.
\newblock In 2022 24th International Conference on Digital Signal Processing
  and its Applications (DSPA).
\newblock \href{https://dx.doi.org/10.1109/DSPA53304.2022.9790744}{Pages 1--5}.
\newblock ~(2022).

\bibitem{higgott_pymatching_2021}
Oscar Higgott.
\newblock ``{PyMatching}: {A} fast implementation of the minimum-weight perfect
  matching decoder''~(2021).
\newblock  url:~\url{http://arxiv.org/abs/2105.13082}.

\bibitem{fowler_minimum_2014}
Austin~G. Fowler.
\newblock ``Minimum weight perfect matching of fault-tolerant topological
  quantum error correction in average \${O}(1)\$ parallel time''~(2014).
\newblock  url:~\url{http://arxiv.org/abs/1307.1740}.

\bibitem{higgott_sparse_2023}
Oscar Higgott and Craig Gidney.
\newblock ``Sparse {Blossom}: correcting a million errors per core second with
  minimum-weight matching''~(2023).
\newblock arXiv:2303.15933 [quant-ph].

\bibitem{mcdermott_quantum--classical_2017}
R.~McDermott, M.~G. Vavilov, B.~L.~T. Plourde, F.~K. Wilhelm, P.~J. Liebermann,
  O.~A. Mukhanov, and T.~A. Ohki.
\newblock ``Quantum--{Classical} {Interface} {Based} on {Single} {Flux}
  {Quantum} {Digital} {Logic}''~(2017).
\newblock arXiv:1710.04645 [quant-ph].

\bibitem{ravi_better_2022}
Gokul~Subramanian Ravi, Jonathan~M. Baker, Arash Fayyazi, Sophia~Fuhui Lin, Ali
  Javadi-Abhari, Massoud Pedram, and Frederic~T. Chong.
\newblock ``Better {Than} {Worst}-{Case} {Decoding} for {Quantum} {Error}
  {Correction}''~(2022).
\newblock arXiv:2208.08547 [quant-ph] version: 2.

\bibitem{krantz_quantum_2019}
P.~Krantz, M.~Kjaergaard, F.~Yan, T.~P. Orlando, S.~Gustavsson, and W.~D.
  Oliver.
\newblock ``A quantum engineer's guide to superconducting qubits''.
\newblock \href{https://dx.doi.org/10.1063/1.5089550}{Applied Physics Reviews
  {\bf 6}, 021318}~(2019).

\bibitem{blais_circuit_2021}
Alexandre Blais, Arne~L. Grimsmo, S.~M. Girvin, and Andreas Wallraff.
\newblock ``Circuit {Quantum} {Electrodynamics}''.
\newblock \href{https://dx.doi.org/10.1103/RevModPhys.93.025005}{Reviews of
  Modern Physics {\bf 93}, 025005}~(2021).

\bibitem{govia2014high}
Luke~CG Govia, Emily~J Pritchett, Canran Xu, BLT Plourde, Maxim~G Vavilov,
  Frank~K Wilhelm, and R~McDermott.
\newblock ``High-fidelity qubit measurement with a microwave-photon counter''.
\newblock \href{https://dx.doi.org/10.1103/PhysRevA.90.062307}{Physical Review
  A {\bf 90}, 062307}~(2014).

\bibitem{opremcak2021high}
Alexander Opremcak, CH~Liu, C~Wilen, K~Okubo, BG~Christensen, D~Sank, TC~White,
  A~Vainsencher, M~Giustina, A~Megrant, et~al.
\newblock ``High-fidelity measurement of a superconducting qubit using an
  on-chip microwave photon counter''.
\newblock \href{https://dx.doi.org/10.1103/PhysRevX.11.011027}{Physical Review
  X {\bf 11}, 011027}~(2021).

\bibitem{di2023discriminating}
L~Di~Palma, A~Miano, P~Mastrovito, D~Massarotti, M~Arzeo, GP~Pepe, F~Tafuri,
  and O~Mukhanov.
\newblock ``Discriminating the phase of a coherent tone with a flux-switchable
  superconducting circuit''.
\newblock \href{https://dx.doi.org/10.1103/PhysRevApplied.19.064025}{Physical
  Review Applied {\bf 19}, 064025}~(2023).

\bibitem{das_afs_2022}
Poulami Das, Christopher~A. Pattison, Srilatha Manne, Douglas~M. Carmean,
  Krysta~M. Svore, Moinuddin Qureshi, and Nicolas Delfosse.
\newblock ``{AFS}: {Accurate}, {Fast}, and {Scalable} {Error}-{Decoding} for
  {Fault}-{Tolerant} {Quantum} {Computers}''.
\newblock In 2022 {IEEE} {International} {Symposium} on {High}-{Performance}
  {Computer} {Architecture} ({HPCA}).
\newblock \href{https://dx.doi.org/10.1109/HPCA53966.2022.00027}{Pages
  259--273}.
\newblock ~(2022).

\end{thebibliography}

\appendix
\begin{figure*}[t!!]
    \centering
    \hspace{-25pt}
    \includegraphics[scale=.3]{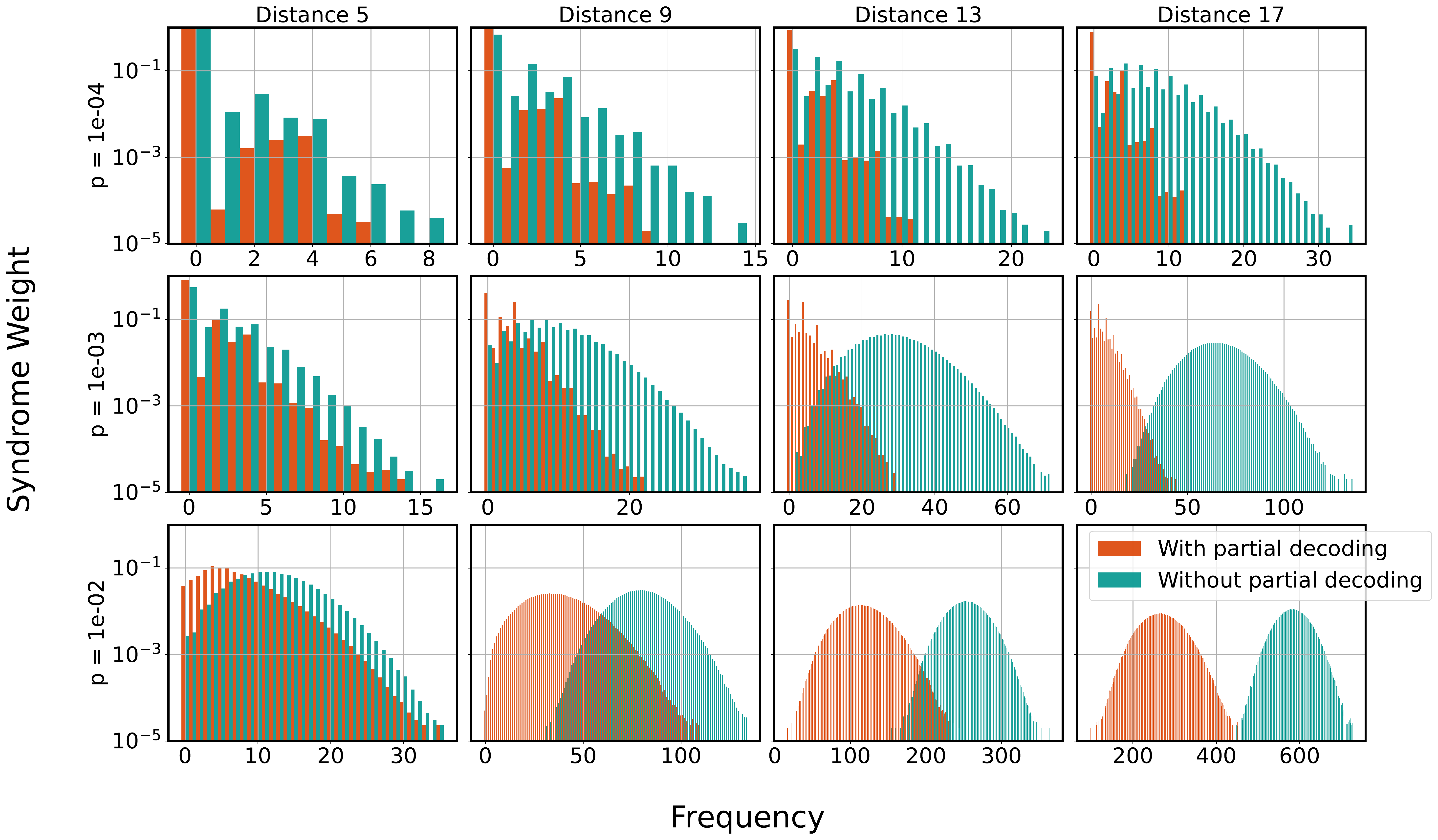}
    \caption{Log-scale histograms of syndrome Hamming weight. The top, middle and bottom row correspond to physical error rate $p=10^{-4}$, $p=10^{-3}$ and $p=10^{-2}$ respectively. The columns, going from left-to-right, correspond to distances $5$, $9$, $13$ and $17$. The green histograms denote the distribution for syndrome weights without partial decoding; the red histograms denote the distribution with partial decoding. }
    \label{fig:app_histograms}
\end{figure*}
\section{Changes in syndrome distributions}\label{app:histograms}
In Fig.~\ref{fig:app_histograms} we show how the distribution of syndromes changes under our partial decoding scheme, for multiple values of the distance $d$ and physical error rate $p$. Each consecutive row corresponds to an increasing physical error rate from $p=10^{-4}$ to $p=10^{-2}$; each consecutive column corresponds to an increasing code distance from $d=5$ to $d=17$. These histograms show the distribution of syndrome weights both without (green) and with (red) partial decoding, on a log scale. Note the unique scale for each x-axis.

At higher physical error rates and distances there are more errors that can happen and hence we observe a wider range of and higher frequency of large syndrome weights. 
This means that the shift in syndrome weight distributions looks more pronounced as the physical error rate and distances are increased. We refer to Fig.~\ref{fig:bandwidth} in Section \ref{sec:sim_results} to show that the syndrome reduction factor is most significant at lower physical error rates.

\section{Belief propagation convergence rate}
\begin{figure}[h]
    \centering
    \includegraphics[width=220pt]{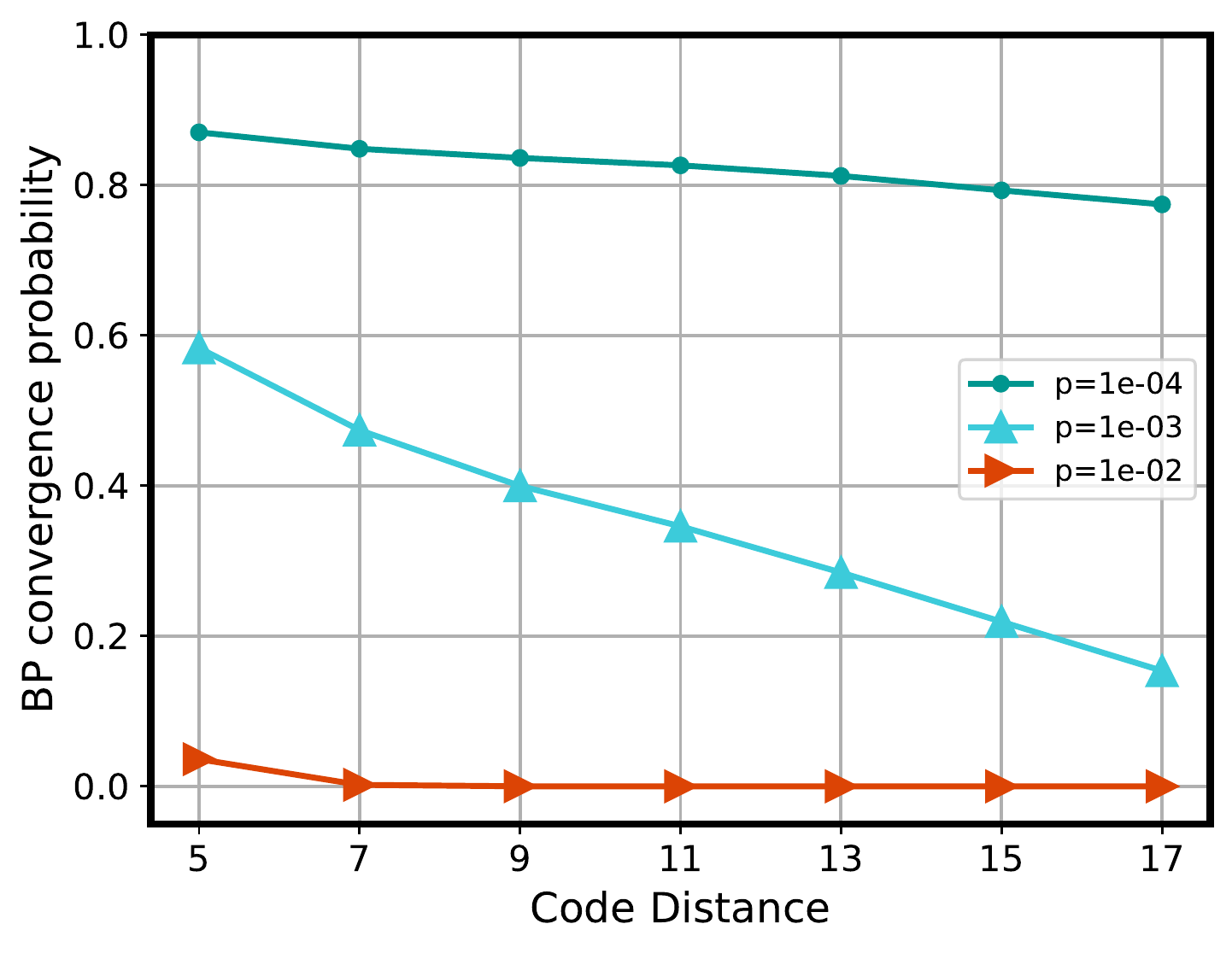}
    \caption{BP convergence probability on non-zero syndromes. Plotted over a range of distances, the lines correspond to physical error rates $1\%$ (red arrowheads), $0.1\%$ (blue triangles) and $0.01\%$ (green dots). BP as partial decoder parameters are $m_{\textrm{iter}}=30$ and $t_{\textrm{BP}}=0.9$.}
    \label{fig:convergence}
\end{figure}

Here we discuss the probability of BP converging on a solution when the syndrome is non-empty. The results are shown in Fig.~\ref{fig:convergence} against code distance, for physical error rates $p=1\%$ (red arrowheads), $p=0.1\%$ (blue triangles) and $p=0.01\%$ (green dots). This plot could equivalently be understood as the probability of not needing a secondary decoder whatsoever. For increasing distances the convergence probability decreases linearly, however with a different gradient for each physical error rate. For $p=1\%$ the probability is not exactly zero, and instead tends to $1$ syndrome in every $50,000$. We note that for $d=17$ and $p=0.01\%$, a regime where fault tolerant quantum computing begins to become feasible, we require a secondary decoder only in $\sim 23\%$ of cases.

\section{Threshold plots} \label{app:threshold_plots}

\begin{figure}[h]
    \centering
    \includegraphics[width=220pt]{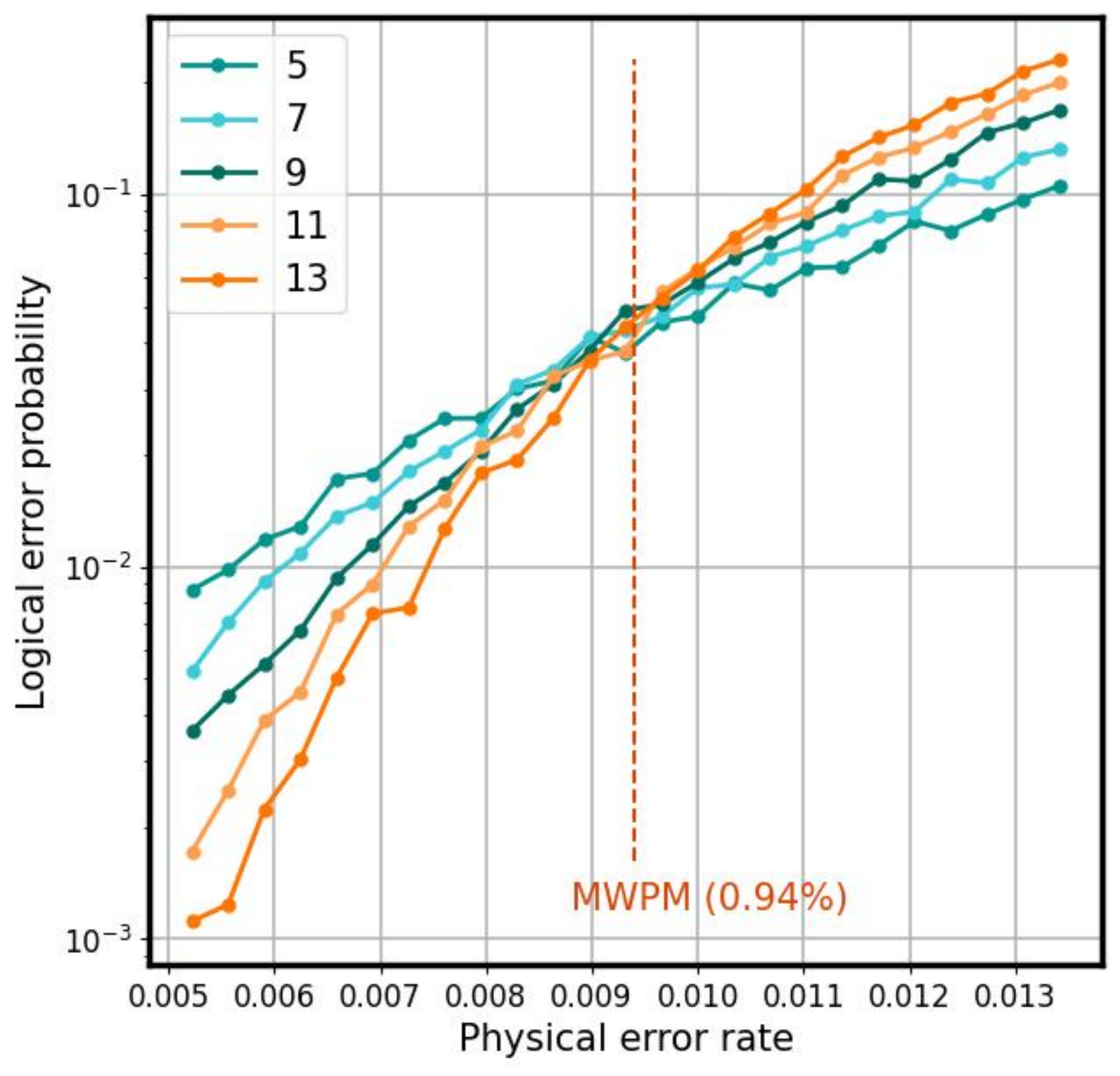}
    \caption{Threshold plot of the surface code $Z$-memory decoded with MWPM under circuit-level noise.}
    \label{fig:accuracy_mwpm}
\end{figure}

\begin{figure}
    \centering
    \includegraphics[width=220pt]{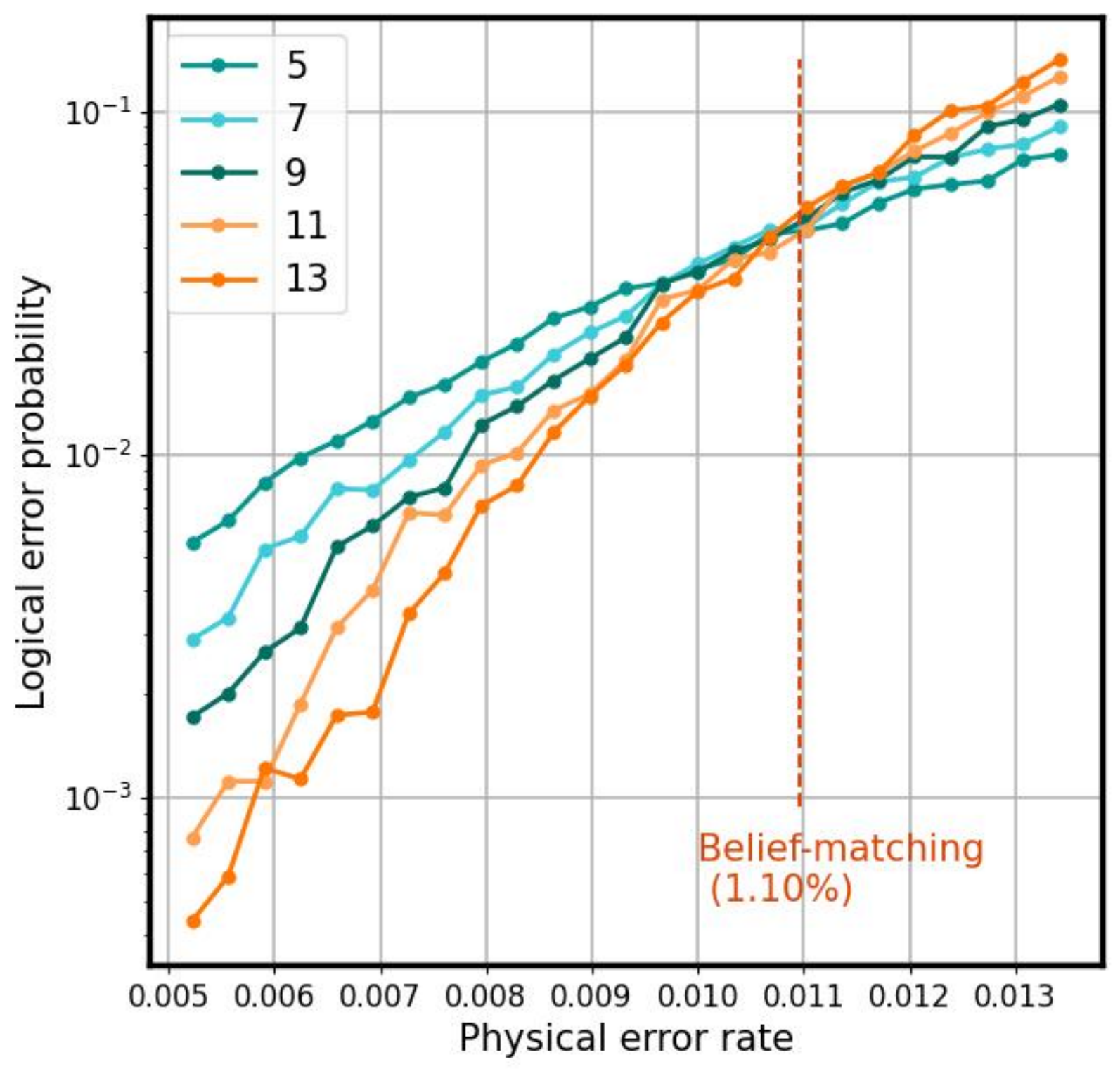}
    \caption{Threshold plot of the surface code $Z$-memory decoded with belief-matching, $m_{\textrm{iter}}=30$ under circuit-level noise.}
    \label{fig:accuracy_belief_matchings}
\end{figure}

Fig.~\ref{fig:accuracy_mwpm} and ~\ref{fig:accuracy_belief_matchings} show the thresholds of surface code $Z$-memory experiment under circuit-level noise decoded using MWPM and belief-matching decoders, respectively.

\end{document}